\documentstyle[12pt]{article}
\input tcilatex
\begin{document}

\begin{center}
{\bf Dynamic quantum theory of large additional dimensions}

\vspace*{1.5cm}

S.N. Andrianov, V.V. Bochkarev

Scientific Center for Gravitational Wave Research Dulkyn, Kazan, Russia
\end{center}

\vspace*{1.5cm}

The Klein-Gordon equation is known in quantum field theory that does not
account the changes of space metrics and changes of particles behavior
connected with it \cite{1}. Such dynamics is describing by Einsteins
equation or Brans-Dicke equation \cite{2}. Wheeler de Witt occupies place of
these equations in quantum theory that is generalization of Klein-Gordon
equation for the case of general relativity theory and is valid for
arbitrary Ryman space \cite{3}. The approach of Wheeler de Witt is applied
to brane theory of Universe in paper \cite{c4}. However, the variation of
bran topology is not accounted in this paper. The variation of space
topology is considered in phenomenological way for quantum theory in paper 
\cite{c5}. In paper \cite{c6}, Wheeler de Witt equation is obtained from
priori taken action for inflating brane. In present paper, we will derive
equation of Wheeler de Witt type in the framework of brane model with the
account of its topology variation in universal space starting from the
well-known conservation laws inside brane.

It is known that the value{\it \ p}$_i${\it p}$^i$ conserves at Lorenz
transform where {\it p}$_i$ is four-dimensional momentum of a particle \cite
{c7}. With that, the following conservation law

\begin{equation}
p_ip^i=0,  \label{1.1}
\end{equation}

is valid for non-massive particle. The three-component vector of momentum
can be written in non-relativistic quantum mechanics in operator form as $%
\overrightarrow{p}=-i\hbar \nabla $. The evident generalization for
four-component case is the value

\begin{eqnarray}
\overline{p}=\left( \frac 1c\widehat{H},\overrightarrow{p}\right) .
\label{e1.2}
\end{eqnarray}

where $\widehat{H}$ is operator of Hamilton since $\widehat{H}\psi =i\hbar 
\frac{\partial \psi }{\partial t}$ according temporal Schr\"odinger
equation. Momentum can be expressed in curvilinear coordinates as covariant
derivative

\begin{equation}
\widetilde{p_i}=-i\hbar \left\{ ...\right\} _{;^{}i}  \label{1.3}
\end{equation}

Lets consider non-mass particle in the three-dimensional Ryman space (brane)
immersed in Deckard space-time of higher dimension. Such brane is can be
shown  by curvilinear axis {\it x} in the moment of time {\it t}. With that,
the momentum of particle $\overline{p}$ is directed along the brane.

The direction of momentum $\widetilde{p}$ does not coincide already with the
direction of curvilinear axis {\it x'} of a brane. Therefore the momentum of
particle directed along the axis {\it x'} is defined due to affine
connection of space-time according rules of covariant differentiation for
fulfilling complete functional variation $\Delta p=p^{\prime }\left(
x^{\prime }\right) -p\left( x^{\prime }\right) +p\left( x^{\prime }\right)
-p\left( x\right) =\delta p+dp$ by the following relation:

\begin{equation}
\overline{p}_i^{\prime }\left( x^{\prime }\right) =\overline{p}_i\left(
x\right) +dp+\delta p_i,  \label{e1.4}
\end{equation}

\begin{equation}
\delta p_i=~\widetilde{p}_k\Gamma _{il}^k\delta x^l.  \label{1.5}
\end{equation}

where $\delta x^l=x^{\prime ~l}-x^l$.

Equation (\ref{e1.4}) can be rewritten in the following form:

\begin{equation}
\overline{p}_i^{\prime }\left( x^{\prime }\right) =\overline{p}_i^{\prime
}\left( x\right) +\delta p_i\left( x,x^{\prime }\right) ,  \label{e1.6}
\end{equation}

where

\begin{equation}
\delta p_i\left( x^{\prime }\right) =\left\{ \Gamma _{il}^k\delta
x^l\right\} _{;~k}^{\prime }  \label{e1.7}
\end{equation}

Substituting expression (\ref{e1.6}) into the formula (\ref{1.1}) we get

\begin{equation}
\ \overline{p}_i\overline{~p}_j+\overline{p}_i\delta p_i+\delta p_i\overline{%
p}_j+\delta p_i\delta p_j=0.  \label{e1.8}
\end{equation}

Being limited by the terms of the first order of magnitude on $\delta p$ and
assuming above mentioned transform to be linear, equation (\ref{e1.8}) can
be transformed using expressions (\ref{e1.2}) and (\ref{e1.7}) to the
following equation on the wave function $\psi $:

\begin{eqnarray}
\nabla _\lambda ^2\psi +\left( R+P_\theta \delta \theta \right) \psi =0.
\label{e1.9}
\end{eqnarray}

where

\begin{eqnarray}
\nabla _\lambda ^2=g^{ij}\left( \frac{\partial ^2}{\partial x^i\partial x^j}%
-\Gamma _{i~j}^k\frac \partial {\partial x^k}\right) ,  \label{e1.10}
\end{eqnarray}

is covariant D'Alabertian,

\begin{eqnarray}
R=g^{ij}R_{ij},  \label{e1.11}
\end{eqnarray}

is scalar curvature, $\delta \theta $ is the variation of additional
coordinate,

\begin{eqnarray}
R_{ij}=\frac{\partial \Gamma _{i~j}^l}{\partial x^k}-\frac{\partial \Gamma
_{i~l}^l}{\partial x^j}+\Gamma _{i~j}^l\Gamma _{l~m}^m-\Gamma _{i~l}^m\Gamma
_{j~m}^l,  \label{e1.12}
\end{eqnarray}

is Richey's tensor,

\begin{eqnarray}
P_\theta =g^{ij}\widehat{P}_{ij\theta },  \label{e1.13}
\end{eqnarray}

is scalar operator,

\begin{eqnarray}
\widehat{P}_{ij\theta }=\frac 12\nabla _i\left( g_{j\theta }R\right)
+R_{i\theta }\frac \partial {\partial x^j},  \label{e1.14}
\end{eqnarray}

The first two terms in the left side of equation (\ref{e1.9}) coincides
formally with the equation of Wheeler de Witt. The last term defined by
operator {\it P}$_\theta $ is the source of curvature at the fluctuation of
additional dimensions.

Let's rewrite equation (\ref{e1.9}) in locally-geodesic coordinate system in
the following form:

\begin{eqnarray}
\left\{ \frac{\partial ^2}{\partial x^2}+\gamma \frac \partial {\partial
x}+a\right\} \psi =\frac 1{c^2}\frac{\partial ^2\psi }{\partial t^2},
\label{e1.15}
\end{eqnarray}

limiting ourselves by one spatial dimension, assuming the absence of affine
connection in time and introducing the notations

\begin{eqnarray}
\gamma =g^{xx}R_{x\theta },  \label{e1.16}
\end{eqnarray}

and

\begin{eqnarray}
a=R+\frac 12\frac \partial {\partial x}\left( g_{x\theta }R\right) .
\label{e1.17}
\end{eqnarray}

Let's look for solution in the form

\begin{eqnarray}
\psi =e^{i\left( kx-\omega ~t\right) }.  \label{e1.18}
\end{eqnarray}

Then we get from (\ref{e1.15}) the characteristic equation

\begin{eqnarray}
k^2-i\gamma k-\left( a+\frac{\omega ^2}{c^2}\right) =0.  \label{e1.19}
\end{eqnarray}

Its physically reasonable solution is

\begin{eqnarray}
k=\sqrt{\frac{\omega ^2}{c^2}+a-\frac{\gamma ^2}4}+i\frac \gamma 2.
\label{e1.20}
\end{eqnarray}

Apparently it is valid when{\it \ a} does not depend on coordinate and time.
When {\it a }= $\gamma =0$, we have the usual dispersion relation {\it a }= $%
\gamma =0$.

When $\left( \frac \omega c\right) \gg a-\left( \frac \gamma 2\right) ^2$ we
can approximately get

\begin{eqnarray}
k=\frac \omega c\left( 1+\frac{c^2}{2\omega ^2}\left( a-\frac{\gamma ^2}%
4\right) \right) +i\frac \gamma 2,  \label{e1.21}
\end{eqnarray}

and for phase velocity of non-massive particles

\begin{eqnarray}
\frac \omega c=\frac c{1+i\frac{\gamma \cdot c}{2\omega }+\frac{c^2}{2\omega
^2}\left( {a-\frac{\gamma ^2}4}\right) }.  \label{e1.22}
\end{eqnarray}

Formula (\ref{e1.22}) shows that effective refractive index related with
space curvature is equal

\begin{eqnarray}
n_{eff}\approx 1+\frac{c^2}{2\omega ^2}\left( {a-\frac{\gamma ^2}4}\right) +i%
\frac{\gamma \cdot c}{2\omega }.  \label{e1.23}
\end{eqnarray}

Thus, the velocity of nonmass particles in universal space can exceed the
velocity of light in plane Deckard space because of drift of particles at
the expansion of Universe. Other consequences of space curvature are the
following two facts that realize when expression (\ref{e1.20}) is valid:

$\bullet $ it is impossible to create particle with kinetic energy less than 
$\hbar \sqrt{\omega ^2+ac^2}$;

$\bullet $ space curvature leads to the frequency shift according the
formula $\omega =\sqrt{\omega _0^2+ac^2}$ that gives the possibility for
verification of developed model when curvature varies at the influence of
gravitational waves.

\end{document}